\documentclass{aa}
\usepackage{graphicx}
\usepackage{epsfig}
\begin{document}

\title{A possible bias on the estimate of $L_{bol}/L_{edd}$ in AGN as a function of luminosity and redshift}

\author{A. Lamastra, G. Matt and G. C. Perola}

\offprints{lamastra@fis.uniroma3.it}

\institute{Dipartimento di Fisica ``E. Amaldi'', Universit\`a degli Studi Roma Tre,
via della Vasca Navale 84, 00146 Roma, Italy \\
Email: lamastra@fis.uniroma3.it 
              }

\date{Received / Accepted }

\authorrunning {A. Lamastra et al.}

\titlerunning {A possible bias on $L_{bol}/L_{edd}$ in AGN}

\abstract{The BH mass (and the related Eddington ratio, $l$ = $L_{bol}/L_{edd}$) in broad line 
AGN is usually evaluated by combining estimates (often indirect) of the BLR radius and of the $FWHM$
 of the broad lines, under the assumption that the BLR clouds are in Keplerian motion around the 
BH. Such an evaluation depends on the geometry of the BLR.  There are two major options for the 
BLR configuration: spherically symmetric or ``flattened''. In the latter case the inclination to 
the line of sight becomes a relevant parameter.
This paper is devoted to evaluate the bias on 
the estimate of the Eddington ratio when a spherical geometry is assumed (more generally when 
inclination effects are ignored), while the actual configuration is ``flattened'', as some
evidence suggests. This is done 
as a function of luminosity and redshift, on the basis of recent results which show the existence 
of a correlation between the fraction of obscured AGN and these two parameters up 
to at least z=2.5 (date at larger redshifts being insufficient.)
The assumed BLR velocity field is akin to the ``generalized thick disk'' proposed by Collin et al. (2006). Assuming an isotropic 
orientation in the sky, the mean value of the bias is calculated as a function of luminosity and 
redshift.
It is demonstrated that, on average, the Eddington ratio obtained assuming a spherical 
geometry is underestimated for high luminosities, and overestimated for low luminosities. This 
bias converges for all luminosities at $z$ about 2.7, while nothing can be said on this 
bias at larger redshifts due to the lack of data. 
The effects of the bias, averaged over the luminosity function of 
broad line AGN, have been calculated. 
The results imply that the bias associated with the 
a-sphericity of the BLR make even worse the discrepancy between the 
observations and the predictions of evolutionary models.

}

\maketitle

\section{Introduction}

The accretion rate and the black hole mass are the two fundamental parameters
in our understanding of the Active Galactic Nuclei (AGN) phenomenon.
Measurements of these two quantities are, unfortunately, not devoid of significant
uncertainties. 

The accretion rate $\dot m$ is derived from the bolometric luminosity, $L_{bol}$, under assumptions on the efficiency $\eta$ for the conversion of gravitational energy, $L_{bol}$ = $\eta$$\dot m$$c^2$. For non-rotating black holes (BH), $\eta$ = 0.057 is generally adopted, assuming that effective (for the observer of the electromagnetic radiation) conversion takes place down to the marginally stable circular orbit at three times the Schwarzschild radius, 3$R_s$; for rotating BH, $\eta$ can reach the maximum value of 0.42. $L_{bol}$ is generally obtained from the luminosity observed in a given band, multiplied by a factor based on the Spectral Energy Distribution (SED) attributed to the specific class the AGN belongs to. This procedure is regarded to be rather safe, but in fact
there are still uncertainties on the luminosity and/or redshift dependence of
the bolometric correction.

For the BH mass, two ``direct'' methods have been followed. The first, applicable 
only to AGN, is the reverberation mapping (RM) method. 
 This method is based 
on the principle (Blandford \& McKee 1982) that the delay in the response of the lines from the Broad
Line Region (BLR) to variations of the continuum is a measure of the size
of this region, $R_{BLR}$. Assuming that the line widths are due to motions
governed by the BH, the combination of $R_{BLR}$ and a velocity derived from
the line profiles yields a ``Keplerian'' estimate of the BH mass. The other
method is based on the fairly strict correlation between the mass of the BH
and properties of the stellar bulge of the host galaxy. This
method, having proved very reliable for a rather large sample 
of galaxies (Gebhardt et al. 2000a, Ferrarese \& Merritt 2000, Tremaine et al. 2002), represents a benchmark for the previous method 
when it can be applied to AGN (Gebhardt et al. 2000b, Ferrarese et al. 2001, Onken et al. 2004). 
The agreement found, although far from perfect (e. g. Collin et al. 2006),
has encouraged the extension to many more AGN (especially the distant and
more luminous ones), for which both the RM and the bulge methods can hardly be
applied, of a ``secondary'' method. The latter is based on the 
estimate of $R_{BLR}$ through an 
empirical correlation between this quantity and the luminosity (see Sect. 2) which
 has emerged from the RM measurements. 

The BH mass can be univocally converted into the Eddington luminosity, $L_{edd}$.
Thus a quantity $l$ can be defined: $l = L_{bol}/L_{edd}$, which, altough it tells us nothing precise about the accretion rate, is of high interest because $L_{edd}$ is a very significant physical limit. This quantity is often referred to as the Eddington ratio.

Several papers have been devoted to explore the behaviour of $l$, in particular
as a function of $L$ and the cosmological epoch, namely the redshift $z$.
Among the uncertainties and the selection effects which may plague the
results, the present paper is devoted to point out and evaluate
a particular bias, linked to the possibility that the
spatial distribution of the BLR clouds is far from spherical, a situation
supported by various lines of evidence. The evaluation is based on 
a recent result on the fraction of AGN which are photoelectrically absorbed
in the X-rays (column density $N_H$ $>$ $10^{22}$ H atoms/cm$^2$, and
Compton thin), which can be summarized as follows. Calling $\xi$ the ratio
of the absorbed ones to the total, it turns out that $\xi$ is a function
of $L_X$ (hence of $L_{bol}$) (Ueda et al. 2003, La Franca et al. 2005)
as well as of $z$ (La Franca et al. 2005). Qualitatively speaking,
in the local Universe this fraction decreases with increasing 
luminosity; as the redshift grows, the anticorrelation remains but
it becomes progressively shallower. If this behaviour is associated with
a luminosity and redshift dependence of the opening angle of the
absorbing matter, within which the BLR can be observed, it should 
introduce a bias on the estimate of the BH mass of broad
line AGN (AGN 1 for short), when this is performed using the RM
method and its ``secondary'' extrapolation. To this effect it is important
to stress (see Fiore et al. 2003,
Perola et al. 2004)
that the value  $N_H$ = $10^{22}$ H atoms/cm$^2$ works as a good
(the exceptions are a minority) discriminant between AGN which are
 optically classified as type 1 and type 2.

The plan of the paper is as follows. In Sect. 2 the results on $l$
from the literature are summarized. In Sect. 3 two lines of evidence,
again from the literature, which
favour a non-spherical distribution of the BLR clouds are briefly 
described. In Sect. 4 the bias associated with the a-sphericity
of the BLR, and its dependence on $L_{bol}$ and $z$,
as it can be predicted on the basis of the abovementioned finding,
is quantified. A discussion follows in Sect. 6.

\section{Estimates of $l$ = $L_{bol}/L_{edd}$: a summary}

If the BLR clouds are orbiting around the BH, the mass of the latter can be estimated as

\begin{equation} \label{Mvirial}
M_{BH}=\frac{R_{BLR} V_{BLR}^{2}}{G}
\end{equation}
where $R_{BLR}$ is the radius of the region and $V_{BLR}$ the velocity of the clouds.
From the application of the RM method (Blandford \& McKee 1982, Peterson 1993) to
a sizeable number of objects, a correlation between $R_{BLR}$ (as estimated
from the ``delay'' in the response of the line chosen for this purpose) and the
luminosity has been found:

\begin{equation}\label{RblrL}
R_{BLR}\propto L^{\alpha}
\end{equation}  
where $\alpha$ $\simeq$ 0.5, and depends somewhat on the band where the luminosity is measured, possibly also 
on which emission line is used (Wandel, Peterson \& Malkan 1999, Kaspi et al. 2000, 2005, Bentz et al. 2006).
Relationship (\ref{RblrL}) is applied to AGN samples which include objects
without a RM estimate.

The most generally used approach is to estimate $V_{BLR}$ from the $FWHM$ of
the line profiles:

\begin{equation}\label{VblrVfwhm}
V_{BLR}=\kappa\times V_{FWHM},
\end{equation} 
where $\kappa$ is a geometrical factor which depends on the shape of the orbits
and on their inclination (Krolik 2001). If the orbits are randomly distributed in 
a spherically simmetric distribution, then (Netzer 1990):

\begin{equation}
 V_{BLR}=\sqrt3/2\times V_{FWHM},
\end{equation}
and eq. (\ref {Mvirial}) becomes

\begin{equation} \label{Msphere}
M_{BH}^{sphere}=\frac{3R_{BLR}V_{FWHM}^{2}}{4G}
\end{equation}

As a general remark, it must be noted that different authors
adopt a different value of $\kappa$, which is then assumed to be {\it constant}. 
Such an assumption is strictly valid only if the width of
the lines, as observed, is independent of the inclination
of the BLR configuration with respect to the line of sight
(namely the configuration is spherically simmetric); or,
with an ``on average'' meaning, if in a sample of objects with
an isotropic distribution in their inclination angle, there
were no limit angle for the BLR to be observable, 
or at least if this angle
were independent of both redshift 
and luminosity.

With this cautionary remark in mind, the results obtained
by various authors (Woo \& Urry 2002, McLure \& Dunlop 2004, Warner
et al. 2004, Kollmeier et al. 2005, Vestergaard 2002, 2004) on different
samples can be summarized in a simple statement: when selection
effects in flux limited samples are taken into account,
there is no evidence that $l$ might depend on $L_{bol}$
or/and on $z$. 
It is hard to evaluate the extent to which this result might 
be influenced by the uncertainties on the extrapolation of 
(\ref {RblrL}) to large
values of $L_{bol}$ and of $z$. In the future it may become feasible
to validate this extrapolation, either directly via the RM tecnique,
or indirectly by comparing the mass estimate with another estimate, obtained 
independently using the ``bulge'' method.
The uncertainty on the bolometric
correction seems less relevant. Although all the abovementioned
authors apply a luminosity independent correction, the
luminosity dependence found by Marconi et al. (2004) is
unlikely to change significantly the results.
Quantitatively, while it is apparent that $l$ is
characterized by a fairly large scatter, and values
larger than unity are also found, its mean value
settles in the range 0.1 to 0.3, depending also
on the value of $\kappa$ adopted.
This applies to $L_{bol}$ up to about 10$^{46}$ erg/s,
and to $z$ values up to 5.

\section{Evidence for a non-spherical BLR}
 
Two lines of evidence in favour of a non-spherical (but axisymmetric)
distribution of the BLR orbits are briefly recalled here. 

The first case concerns AGN which are also radiogalaxies
or quasars. Wills \& Browne (1986) found a highly significant
anticorrelation between the $FWHM$ of broad H$\beta$ lines and 
the radio parameter $R$ defined as the ratio between the flux densities of the core 
and of the radio lobes. In the relativistic beaming models
of radiogalaxies (Blandford \& Rees 1978, Orr \& Browne 1982, 
Hough \& Readhead 1989) the difference between core dominated (large $R$)
and lobe dominated (small $R$) sources is attributed to a difference in orientation:
the first are those viewed close to the beam axis, the second
those viewed at larger angles. Thus $R$ can be regarded as an indicator
of the orientation angle of the beam, and if the beam coincides
with the axis of simmetry of the BLR, the anticorrelation just
mentioned can be read as indicating that the orbits of the clouds
are predominantly confined to a plane perpendicular to that axis.
This analysis has been refined and the conclusion confirmed
by Wills \& Brotherton (1995).

There is no direct evidence that the same picture applies to
the radio quiet AGN as well, but several arguments 
have been proposed in favour of a flattened, rather than a spherical, BLR 
(Rudge \& Raine 1999, McLure \& Dunlop 2002, Jarvis \& McLure 2006).

Collin et al. (2006) made the second case on the basis of some
significant discrepancies between BH masses estimated with
the RM method and those estimated with the ``bulge'' method
in the same objects. They show that such discrepancies 
are most marked for the objects with the narrowest lines, which
are likely the objects with their flattened BLR seen almost pole-on.
 This conclusion is not substantially affected by the use
of the second moment of the line profile, instead of its $FWHM$,
adopted by Collin et al. (2006).

\section{A bias on $l$ due to a flattened BLR}

Following Collin et al. (2006), a simple parameterization (which they
refer to as a ``generalized thick disk'') will be adopted, 

\begin{equation}\label{Vdisc}
 V_{BLR}=\frac{V_{FWHM}}{2(a^{2}+\sin^{2}i)^{1/2}},
\end{equation}
where $i$ is the inclination angle of the disk axis relative 
to the line of sight. When (\ref {Vdisc}) is inserted 
into eq. (\ref {Mvirial}), an estimate of the BH mass is obtained, called here $M_{BH}^{disk}$:

\begin{equation} \label{Mdisc}
 M_{BH}^{disk}=\frac{R_{BLR}V_{FWHM}^{2}}{4G(a^{2}+\sin^{2}i)}.
 \end{equation}

\begin{figure}[h]
\begin{center}
\includegraphics[width=8 cm]{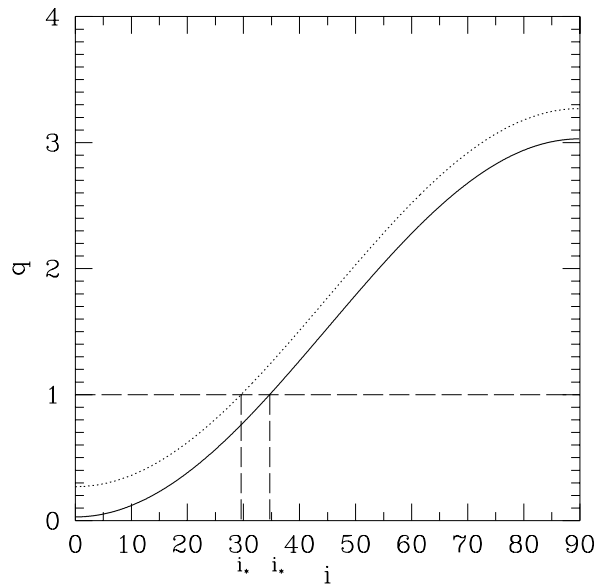}
\caption{The ratio $q = M_{BH}^{sphere}/M_{BH}^{disk}$ as a function of the
inclination angle $i$, assuming $a$=0.1 (solid line) and $a$=0.3 (dotted line).}
\label{Mbhspheredisc}
\end{center} 
\end{figure}

Comparing eq. (\ref {Msphere}) with eq. (\ref {Mdisc}), it 
is apparent that at the angle $i_{\ast}$=$\arcsin \sqrt{(1/3-a^{2})}$
the two estimates give the same value ($i_{\ast}$ $\simeq$ 35$^{\circ}$ for $a$ = 0.1 and $i_{\ast}$ $\simeq$ 30$^{\circ}$ for $a$ = 0.3). 
It is therefore convenient
to introduce the ratio

\begin{equation} \label{qi}
 q=\frac{M_{BH}^{sphere}}{M_{BH}^{disk}}=3(a^{2}+\sin^{2}i),
 \end{equation}
which is illustrated in Fig. \ref {Mbhspheredisc} for two
values of $a$, namely 0.1 and 0.3. The quantity $q$
decreases rather quickly with $i$, when $i < i_{\ast}$, and 
is fairly sensitive to the value of $a$. For instance, for $i = 10^{\circ}$,
$q$ is equal to 0.12 ($a$ = 0.1), or to 0.36 ($a$ = 0.3). When $i > i_{\ast}$, $q$ increases up to a maximum
of about 3 for both values of $a$.
 
As already emphasized by Collin et al. (2006), for a {\it single} object without
an independent observational constraint on its inclination angle,
the estimate of the BH mass using eq.(\ref {Msphere}) could be in 
error by up to a factor of 100. However, the interest here is on systematic
errors on {\it large} samples. 

If the orientation in the sky of the disk axis is assumed isotropic, namely

\begin{equation}\label{dndi}
\frac{\mathrm{d}N}{\mathrm{d}i}=\sin i,
\end{equation}
it immediately follows that, if there were no limits to the angle
at which the BLR can be observed, the
mean value of $q$ would be 2 ($a$=0.1) or 2.3 ($a$=0.3). If a spherical distribution
is adopted (similarly for a $\kappa$ value of 1), $M_{BH}$ is 
significantly overestimated, and hence $l$ underestimated. This is relevant
in itself, but before discussing in this respect the results
reported in Sect. 2, one must deal with the fact that in the AGN unification model the
type 1 objects can be found only within a limiting angle, say $i_{0}$, as
pointed out by Collin et al. (2006). In addition this angle, according to the 
findings by Ueda et al. (2003) and La Franca et al. (2005),
is a function of both luminosity and redshift.

\begin{figure}[h]
\begin{center}
\includegraphics[width=8 cm]{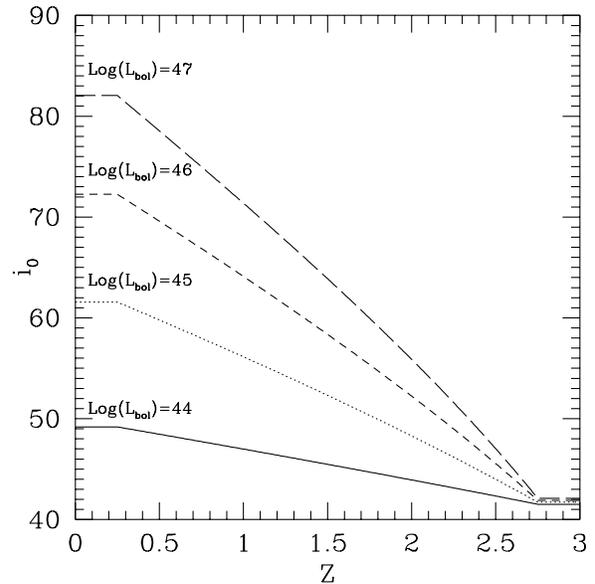}
\caption{The opening angle $i_0$, inferred from the fraction $\xi$ in La Franca et al. (2005),
as a function of $z$ for different values of the AGN bolometric luminosity. 
}
\label{angleZLbol}
\end{center} 
\end{figure}

With reference to the results of La Franca et al. (2005) on the fraction $\xi$ as a function of
$L_X$ and $z$, obtained taking into
account the selection effects on the samples used (illustrated in their Figure 11), one can immediately 
derive an opening angle $i_0$:\begin{equation}\label{ioxi}
\cos i_{0}=\xi(L_{X},z)
\end{equation} 
which discriminates type 1 AGN (where the BLR is visible) 
from type 2 AGN (where it is not).
The function $\xi$, in La Franca et al. (2005), holds in the ranges 0.25$\leq z \leq$2.75 and 42.5$\leq \log\!L_X \leq$45.5. At redshifts and luminosities outside this ranges, $\xi$ is kept costant and equal to the values obtained at the limits of the intervals.
The outcome is illustrated in Fig. \ref {angleZLbol}, where $L_X$ has been converted to $L_{bol}$
using the (luminosity dependent) bolometric correction 
given by Marconi et al. (2004). Each line refers to a luminosity interval
of $\pm 0.5$ dex.

It must be stressed that La Franca et al. (2005) used a simple functional
form to describe the quantity $\xi$, which does probably represent fairly
well the real situation up to $z$ around 2. However, it cannot be extrapolated to
larger values without a sizeable change in slope, which however could not
be properly estimated with the data available. The authors preferred to
introduce a sort of ``saturation'' value, which explains why the curves 
for the four different values of $L_{bol}$ intersects at $z$ = 2.7, with
$i_0$ = 41$^{\circ}$, which corresponds to the ``saturation'' value of $\xi$ = 75\%.

\begin{figure*}[ht]
\hbox{
\includegraphics[width=7 cm]{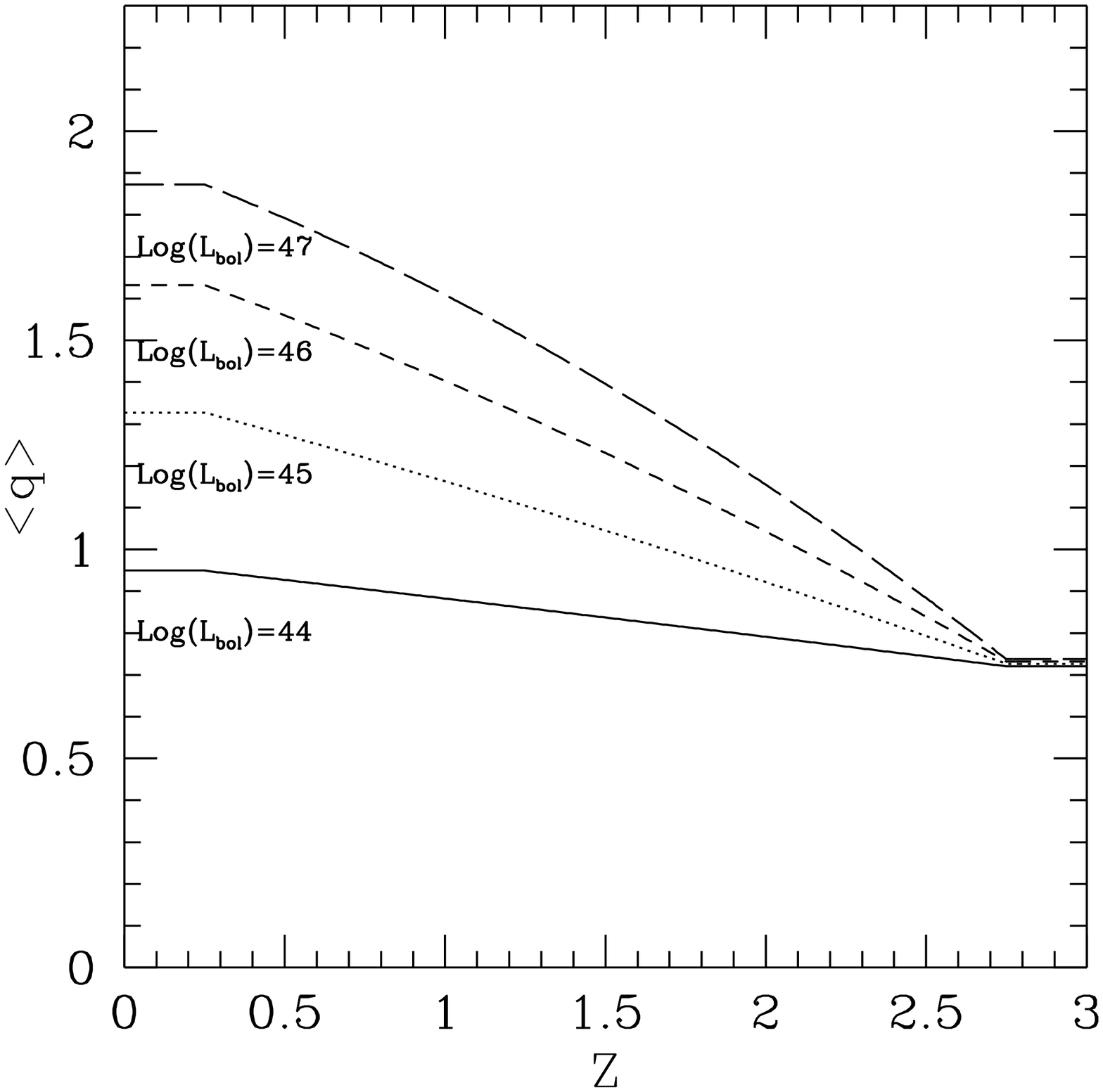}
\hspace{0.02cm}
\includegraphics[width=7 cm]{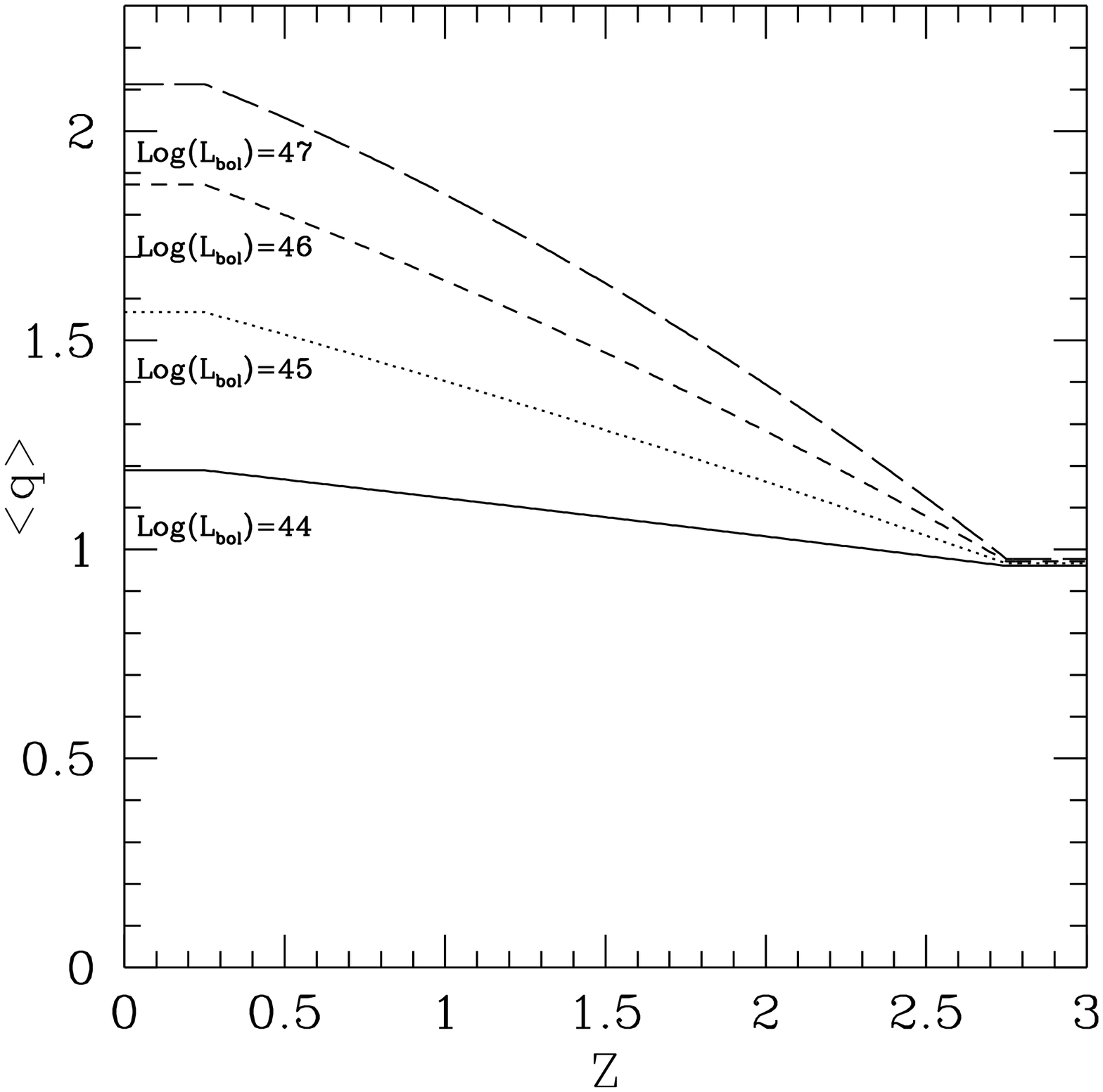}
}
\caption{Mean value of $q$, $<q>$, as a function of redshift for different values of the bolometric luminosity; $a$=0.1 (left) and $a$=0.3 (right).}
\label{qmedioZLbol}
\end{figure*}

At this point, assuming that the disk axis and the obscuring matter axis are coincident, 
the mean value of $q$, as a function of luminosity and redshift, can be derived using eq. (\ref {dndi}):

\begin{equation} \label{qmedio}
<q>=\frac{\int_{0}^{i_{0}(L_X,z)}q\, \sin i\, \mathrm{d}i}{\int_{0}^{i_{0}(L_X,z)}\sin i\, \mathrm{d}i},
\end{equation}
where $i_0$(L$_X$,z) is given by eq. (\ref {ioxi}). The results are illustrated in Fig. \ref {qmedioZLbol} for the values
$a$ = 0.1 and 0.3.

%

\section{Discussion}

The curves in Fig. \ref {qmedioZLbol} show that the mean value of the
bias, $<$q$>$, is typically greater than 1 and can be as large as about 2.
Furthermore, this bias increases with the luminosity,  converging
for all luminosities towards a value of order unity at $z$
larger than 2. As noted previously, the estimate of the opening angle $i_0$, hence of  $<$q$>$, ceases to be 
reliable in this regime of the cosmic time, because the lack of
data prevented La Franca et al. (2005) from evaluating the actual evolution
of $\xi$ further back in time. The ``saturation'' effect adopted in $\xi$
corresponds to an angle $\simeq 40^{\circ}$,  such that, according to eq. (\ref {qmedio}), $<$q$>$ is $\simeq 1$ for 
$a$=0.3, or $\simeq 0.7$ for $a$=0.1. Since the weight of the solid angle within this value of $i_0$ is relatively modest, the mean value of q will result significantly lower than unity only for $i_0$ much lower than $i_{\ast}$. If, for instance, $\xi$ were to reach 90\% (that is $i_0$ = 26$^{\circ}$)
at values of $z$ much greater than 2, than  $<$q$>$ would be 0.32 ($a$=0.1) or 0.56 ($a$=0.3). 

We also calculated the rms of the $q$ distribution and found values of 0.9, 0.6 and 0.2 for $a$=0.1
and $i_0$=90$^{\circ}$, 60$^{\circ}$, 30$^{\circ}$, respectively (the corresponding values of 
$<$q$>$ are 2.0, 1.3 and 0.4). Similar values are found for $a$=0.3. The expected scatter in masses is thus
reassuringly smaller than that found in the Black Hole mass (estimated from the reverberation mapping) 
vs. bulge dispersion velocity relationship (Onken et al. 2004).

Turning to the parameter $l$, this quantity is {\it underestimated} when $<$q$>$ $>$ 1, if $M_{BH}$
is ``measured'' assuming a spherical distribution of the BLR, the other way round if $<$q$>$ were less than unity. 

A possible way to illustrate the effects of this bias consists in calculating its value
averaged over the entire Luminosity Function. La Franca et al. (2005) give both the Luminosity
Function (LF) and the $\xi$ which best fit the data (fit \#4 in their Table 2), after taking into account the selection
effects (in the X-ray and optical bands) for the samples used. The LF behaviour with cosmic time follows a Luminosity Dependent Density
Evolution. One can therefore combine the LF and $\xi$ to obtain, for a given
$z$, the average bias, $<$q$>$$_{L}$, over the entire luminosity range:

\begin{equation} \label{qmedioL}
 <q>_{L}=\frac{\int_{\log\!L_{X1}}^{\log\!L_{X2}}\frac{\mathrm{d}\Phi_{1}(L_{x},z)}{\mathrm{d}\!\log\!L_{X}}\, \mathrm{d}\!\log\!L_{X}\int_{0}^{i_{0}(L_{X},z)}q\, \sin i\, \mathrm{d}i}
{\int_{\log\!L_{X1}}^{\log\!L_{X2}}\frac{\mathrm{d}\Phi_{1}(L_{x},z)}{\mathrm{d}\!\log\!L_{X}}\, \mathrm{d}\!\log\!L_{X}\int_{0}^{i_{0}(L_{X},z)}\sin i\, \mathrm{d}i},
 \end{equation}
where $\Phi_{1}$ is the luminosity function of unabsorbed AGN (which depends on $\xi$) and the parameter $a$ is included in the function $q$ (eq. (\ref {qi})).
This result should, by construction, be free from selection effects. For a choice of $\log\!L_{X1}$ = 42 and $\log\!L_{X2}$ = 48, $<$q$>$$_{L}$  is illustrated in the left panel of  Fig. \ref{qmedioL} (the corresponding values of $\log\!L_{bol}$, obtained applying the luminosity dependent bolometric correction of Marconi et al. (2004), are: $\log\!L_{bol1}$ = 43.030 and $\log\!L_{bol2}$ = 50.937).

\begin{figure*}[ht]
\hbox{
\includegraphics[width=7 cm]{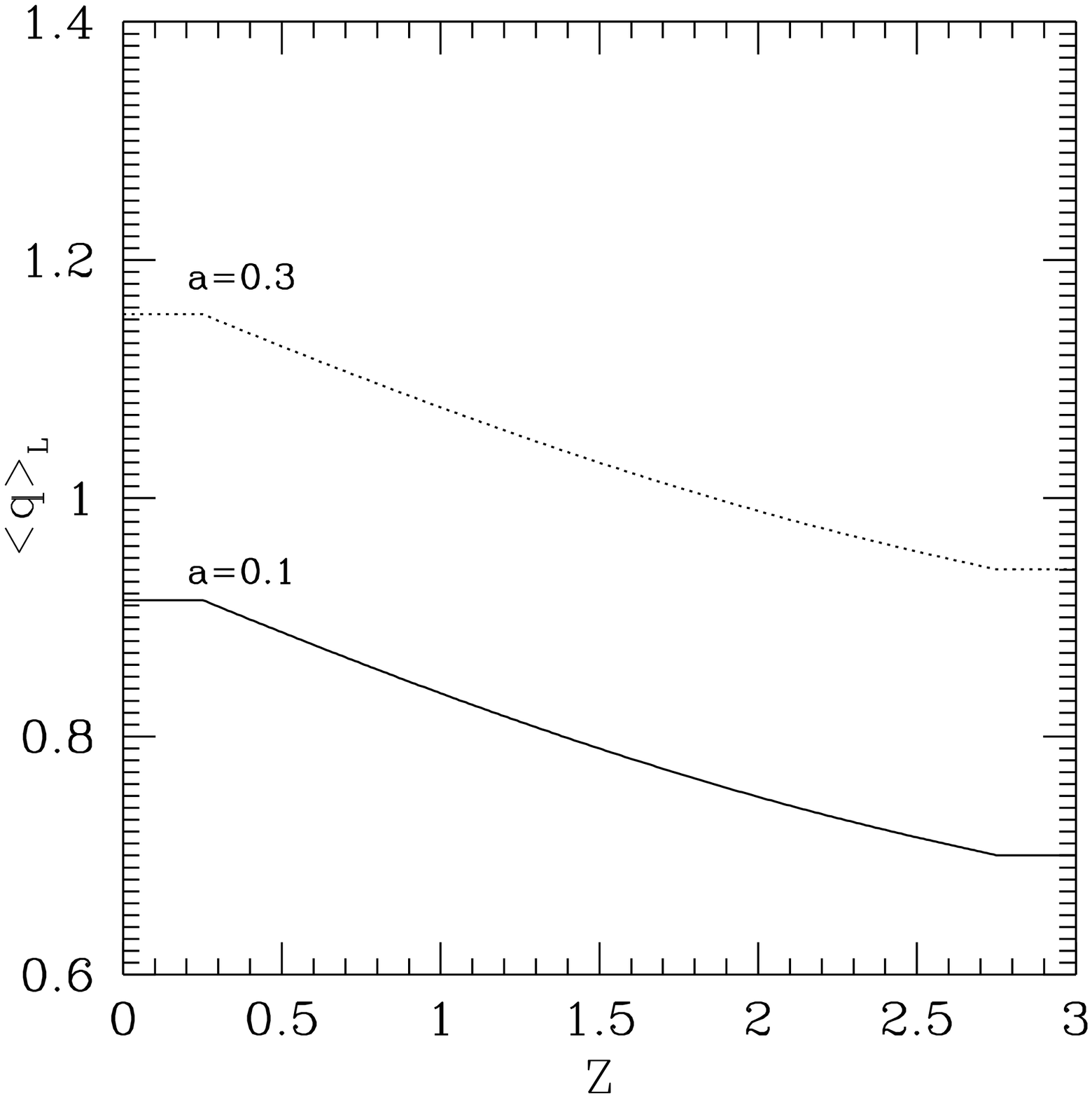}
\hspace{0.02cm}
\includegraphics[width=7 cm]{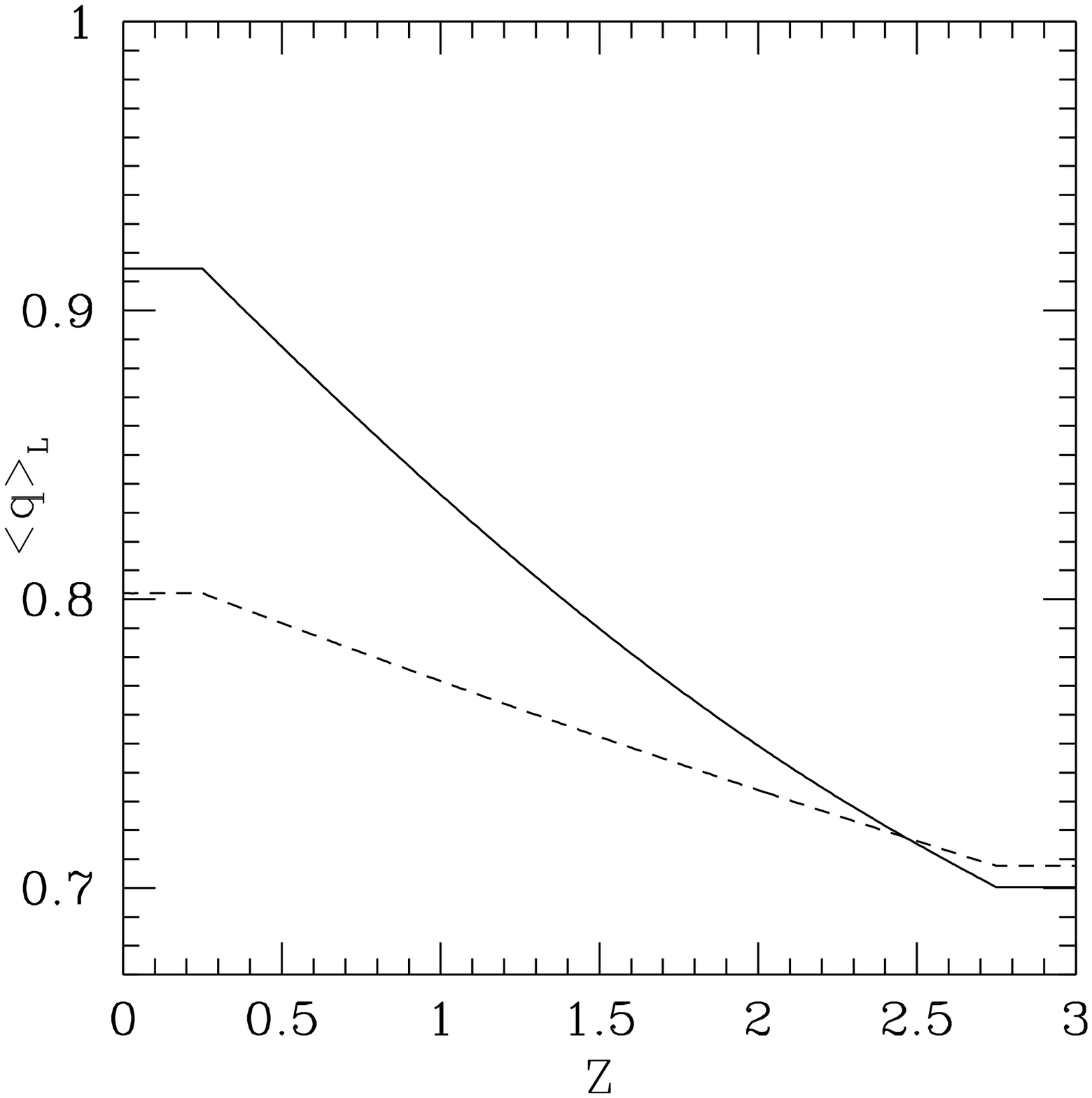}
}
\caption{Left: mean value of q, $<$q$>$$_{L}$, between 42$<\log\!L_{x}<$48 as a function of z, for a=0.1 (solid line) and a=0.3 (dotted line). Right: mean value of q, $<$q$>$$_{L}$, between 41$<\log\!L_{x}<$48 (dashed line) and  42$<\log\!L_{x}<$48 (solid line) as a function of z, for a=0.1.}
\label{qmedioL}
\end{figure*}


To illustrate the effect arising from the inclusion of the numerous AGN in the decade around  $\log\!L_{X}$ = 41 ($\log\!L_{bol}$ = 41.893), in the right panel of Fig. \ref {qmedioL} (and only for the case $a$=0.1) 
$<$q$>$$_{L}$ is compared with the result shown in the left panel of the same figure. The effect 
is that, the higher is the luminosity, the steeper is the dependence of the mean bias $<$q$>$$_{L}$ on $z$.


A direct application of our results to the samples used by the various authors, quoted in Sect. 2, is not straightforward, because this would require proper evaluation of their specific observational selection effects, a hard task which goes beyond the aims of this paper. However, some general remarks can be made. The results shown in Fig. \ref {qmedioL} imply that, if $l$ were intrinsically constant with $z$, and if its ``mean'' value were calculated with the spherical approximation (or, more generally, with a constant value of $\kappa$ in eq. (\ref {VblrVfwhm})) then one should observe an increase of $l$ with $z$. Since the observational results indicate instead a constant value, the bias discussed in this paper implies that the actual value of $l$ \textit{decreases} with $z$. 

In semianalytical models which link the evolution of the galaxies in the hierarchical clustering scenario with the quasar evolution (e.g. Menci et al. 2003, 2004), the black hole accretion is triggered by galaxies encounters. In this scenario, at high $z$ the protogalaxies grow rapidly by hierarchical merging, meanwhile much cold gas is imported and also destabilized, so that the black holes are fueled at their Eddington rates. At lower $z$ the accretion rate of cold gas onto the central black hole diminish due to the combined effects of the decrease of the galaxies merging and encounter rates and the decrease of the amount of galactic cold gas, which was already converted into stars or accreted onto the black hole. This model predicts an average Eddington ratio dropping from $l$ $\simeq$ 1 at $z$ $\simeq$ 2.5 to $l$ $\simeq$0.01 at $z$ $\simeq$0 (Menci et al. 2003). 
Our results imply that the bias associated with the 
a-sphericity of the BLR make even worse the discrepancy between the observations and the predictions of the models.

\acknowledgements{The authors thank Fabio La Franca for his help with the luminosity function, and INAF-ASI for the grant I/023/05/0.}


\begin{thebibliography}{}

\bibitem{} Bentz, M. C.; Peterson, B. M.; Pogge, R. W.; Vestergaard, M.; Onken, C. A., 2006, ApJ, 644, 133

\bibitem{} Blandford, R.D., $\&$ McKee, C.F., 1982, ApJ, 255, 419 

\bibitem{} Blandford, R.D., $\&$ Rees, M., 1978, in Proc. Pitt. Conf. on BL Lac objects, ed. Wolfe A. M. (Pitt. Univ. of Pitt), 328 

\bibitem{} Collin, S., Kawaguchi, T., Peterson, B. M., Vestergaard, M., 2006 A$\&$A, in press (astro-ph/0603460)

\bibitem{} Ferrarese, L., Merritt, D., 2000, Apj, 539, L9

\bibitem{} Ferrarese, L., Pogge, R. W.. Peterson, B. M., Merritt, D., Wandel, A., Joseph, C. L., 2001, ApJ, 555, L79F  

\bibitem{} Fiore, F., Brusa, M., Cocchia, F., et al. 2003, A$\&$A, 409, 79

\bibitem{} Gebhardt, K., et al., 2000a, ApJ, 539, L13

\bibitem{} Gebhardt, K., et al., 2000b, ApJ, 543, L5

\bibitem{} Hough, D. H., $\&$ Readhead, A. C. S., 1989, AJ, 98, 1208

\bibitem{} Jarvis, M. J., $\&$ McLure, R. J., MNRAS 369, 182

\bibitem{} Kaspi, S., Maoz, D., Netzer, H., Peterson, B. M., Vestergaard, M.,
 $\&$ Jannuzi, B. T. 2005, ApJ, 629, 61

\bibitem{} Kaspi, S., Smith, P.~S., Netzer, H., Maoz, D., Jannuzi, B.~T., $\&$ Giveon, U.\ 2000, ApJ, 533, 631 

\bibitem{} Kollmeier, J. A., Onken, C. A., Kochanek, C. S., Gould, A., Weinberg, D. H., Dietrich, M., Cool, R., Dey, A., Eisenstein, D. J., Jannuzi, B. T., Le Floc'h, E., Stern, D., 2005 ApJ, in press (astro-ph/0508657)

\bibitem{} Krolik, J.~H.\ 2001, ApJ, 551, 72

\bibitem{} La Franca, F., Fiore, F., Comastri, A., et al., 2005, ApJ, 635, 864L.  

\bibitem{} Marconi, A., Risaliti, G., Gilli, R., 2004, MNRAS, 351, 169

\bibitem{} McLure, R. J., Dunlop, J. S., 2002, MNRAS, 331, 795

\bibitem{} McLure, R. J., Dunlop, J. S., 2004, MNRAS, 352, 1390M	
	
\bibitem{} Menci, N., Cavaliere, A., Fontana, A., Giallongo, E., Poli, F., Vittorini, V., 2003, ApJ, 587, L63

\bibitem{} Menci, N., Fiore, F., Perola, G. C., Cavaliere, A., 2004, ApJ, 606, 58

\bibitem{} Netzer, H., 1990, in Active Galactic Nuclei, ed.\ T.~J.-L. Courvoisier \& Mayor, M. (Springer Verlag: Berlin), p.\ 137

\bibitem{} Onken, C. A., Ferrarese, L., Merritt, D., Peterson, B. M., Pogge, R. W., Vestergaard, M., $\&$ Wandel, A., 2004, ApJ, 615, 645

\bibitem{} Orr, M. J. L., Browne, I. W. A., 1986, MNRAS, 200, 1067O 

\bibitem{} Perola, G. C., et al., 2004, A$\&$A, 421, 491

\bibitem{} Peterson, B.M. 1993, PASP, 105, 247

\bibitem{} Rudge, C. M., Raine, D. J., 1998, MNRAS, 297, L1

\bibitem{} Tremaine, S., et al., 2002, ApJ, 574, 740

\bibitem{} Ueda, Y, Akiyama, M., Ohta, K., Miyaji, T., 2003, ApJ, 598, 886

\bibitem{} Vestergaard, M., 2002, ApJ, 571, 733

\bibitem{} Vestergaard, M., 2004, ApJ, 601, 676

\bibitem{} Wandel, A., Peterson, B.~M., $\&$ Malkan, M.~A.\ 1999, ApJ, 526, 579 

\bibitem{} Warner, C., Hamann, F., Dietrich, M., 2004, ApJ, 608, 136W	

\bibitem{} Wills, B. J., Browne, I. W. A., 1986, ApJ, 302, 56W

\bibitem{} Wills, B. J., Brotherton, M. S., 1995, ApJ, 448L, 81W

\bibitem{} Woo, J., Urry, C. M., 2002, ApJ, 579, 530W



\end{thebibliography}
\end{document}